\documentclass[12pt]{iopart}
\usepackage{graphicx}% Include figure files
\usepackage{iopams}
\usepackage{setstack}
\usepackage{bm}% bold math
\input{epsf}

\def\beq{\begin{equation}}
\def\eeq{\end{equation}}
\def\ba{\begin{eqnarray}}
\def\ea{\end{eqnarray}}
\def\agame{\mathcal{A}}
\def\bgame{\mathcal{B}}
\def\cgame{\mathcal{A}\star\mathcal{B}}

\def\b0{\arrowvert_{\beta=0}}

\begin{document}
\title{Recycling Parrondo games}

\vskip4mm                      

\author{Roberto Artuso\dag\footnote[1]{also at Istituto Nazionale Fisica Nucleare, Sezione di Milano, Via Celoria 16, 20133 Milano, Italy, and CNR-INFM, Sezione di Como}, Lucia Cavallasca\dag and Giampaolo Cristadoro\ddag}
\address{\dag\ Center for Nonlinear and Complex Systems and
Dipartimento di Fisica e Matematica, Universit\`a
dell'Insubria, Via Valleggio 11, 22100 Como, Italy}
\address{\ddag\  Max Planck Institute for the Physics of Complex Systems, N\"{o}thnitzer Str. 38, D-01187 Dresden, Germany}
%\eads{\mailto{Roberto.Artuso@uninsubria.it}, \mailto{Giampaolo.Cristadoro@uninsubria.it}
\begin{abstract}
We consider a deterministic realization of Parrondo games, and use periodic orbit theory to analyze their asymptotic behavior.
\end{abstract}
\submitto{\JPA}
\pacs{05.45.-a }
\maketitle
%\begin{multicols}{2}
%\narrowtext

\section{Introduction}
Directed transport has received a remarkable attention for a number of years now, since it has been recognized as a crucial issue in a number of physical and biological contexts (see the review \cite{reimann} and references therein). A striking phenomenon in this context is {\em ratchets} behavior, where currents may flow in a counter intuitive direction. Parrondo games (see \cite{Parr1} and references therein) offer a remarkable and simple illustration of these subtle transport properties: a random combination of two losing games may result in a winning strategy. In this paper we present a deterministic analogue of Parrondo games, in the form of a (piecewise linear), periodic map on the real line, and study transport properties by means of periodic orbit expansions \cite{Webbook}: in this simple setting they allow to perform analytic evaluations of the relevant quantities, while providing a highly effective perturbative technique to get accurate estimates in more general cases (for instance by considering nonlinear mappings). The paper is organized as follows: in section $2$ we construct a deterministic realization of Parrondo games in the form of a one-dimensional mapping on the real line, in section $3$ we briefly review how to study transport properties of deterministic systems by periodic orbits expansions, in section $4$ we apply such a theory to Parrondo mappings and discuss some features of our findings, and in section $5$ we give our conclusions, and possible future developments of the present work.
\section{One-dimensional Parrondo mappings}
Let us briefly recall the simplest implementation of Parrondo games: we start from two simple games $\mathcal{A}$ and $\mathcal{B}$: $\mathcal{A}$ is a simple coin tossing game with winning probability {\em p} and losing probability $1-\hbox{\textit{p}}$: each time the game is played the player toss the coin: if he/she wins then the capital (an integer value, which we denote by $X$) is increased by one unit, otherwise it is decreased by the same amount. Game $\mathcal{B}$ is a little bit more complicated, as it requires two ``coins'', selected upon inspection of the present value of the capital: if $X=0{\,}_{mod\,\,M}$ winning/losing probability is $\hbox{\textit{p}}_1$/$1-\hbox{\textit{p}}_1$, while in the opposite case ($X\neq 0{\,}_{mod\,\,M}$) the winning/losing probability is $\hbox{\textit{p}}_2$/$1-\hbox{\textit{p}}_2$: $M$ is a fixed integer. Now suppose that at any integer time step $n$ (starting from zero) a game is played: $X(n)$ with represent the instantaneous value of the capital. It turns out that fine tuning of the parameters leads to a paradoxical behavior: namely take $M=3$, $p=1/2 -\epsilon$, $p_1=1/10-\epsilon$, $p_2=3/4-\epsilon$, for a sufficiently small value of $\epsilon$: both games are in this case slightly unfair: if the player keeps on playing $\mathcal{A}$ or $\mathcal{B}$ the capital will drop down linearly, while playing $\mathcal{A}$ or $\mathcal{B}$ in random order (for instance with probability $1/2$) results in a winning strategy, where on the average the capital increases linearly with time! The paradox can be explained by a Markov chain analysis \cite{Parr2}. 

\begin{figure}[h!]
%\label{normvsint}
\centerline{\epsfxsize=8.cm \epsfbox{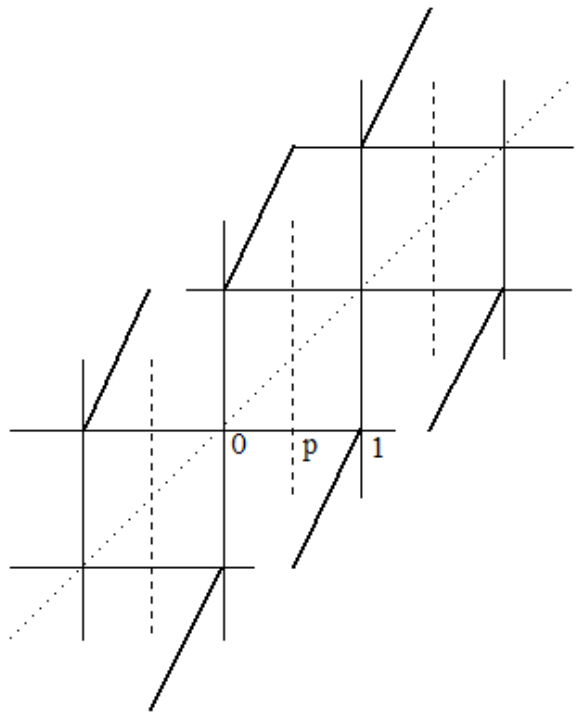}}
\vskip4mm
\caption{\label{Areal}The map corresponding to game $\mathcal{A}$}
\end{figure}

\begin{figure}[h!]
%\label{normvsint}
\centerline{\epsfxsize=6.cm \epsfbox{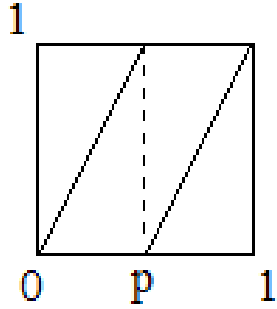}}
\vskip4mm
\caption{\label{Atorus}The torus map corresponding to game $\mathcal{A}$}
\end{figure}

We now map the problem to a dynamical setting: that is we want to introduce a map $F_P$ on the real line such that $X(n+1)=F_P(X(n))$ where ``transition probabilities'' are a reflection of the game rules. Let's start from game $\mathcal{A}$: the dynamical
realization is easily accomplished once we define $F_P$ on the unit interval as follows:
\beq
F_{P\agame}(x)\,=\left\{
\begin{array}{ll}
\frac{1}{p} x+1 & \mbox{  } x\in[0,p) \nonumber\\
\frac{1}{1-p} x - \frac{1}{1-p} & \mbox{  } x\in[p,1) \nonumber
\end{array}
\right.
\label{FPA}
\eeq
and then extend (\ref{FPA}) on the real line by translation symmetry
\beq
F_{P\agame}(x+n)\,=\,F_P(x)+n\qquad \quad n\in\mathbb{Z}
\label{tsA}
\eeq
(see Fig.(\ref{Areal})).
The left branches of the map correspond to a winning event (capital is increased by
one unit, strictly speaking capital corresponds to the integer part of the dynamical variable here) while right branches account for losing events. In the theory we will employ later on together with the map on the real line we also need to consider a corresponding torus map $\hat{F_P}(\vartheta)$ on the circle, that is simply defined as (see Fig. (\ref{Atorus}))
\beq
\hat{F}_{P\agame}(\vartheta)\,=\,F_P(\vartheta){\,}_{mod \, 1}
\label{FPAtorus}
\eeq

\begin{figure}[h!]
%\label{normvsint}
\centerline{\epsfxsize=10.cm \epsfbox{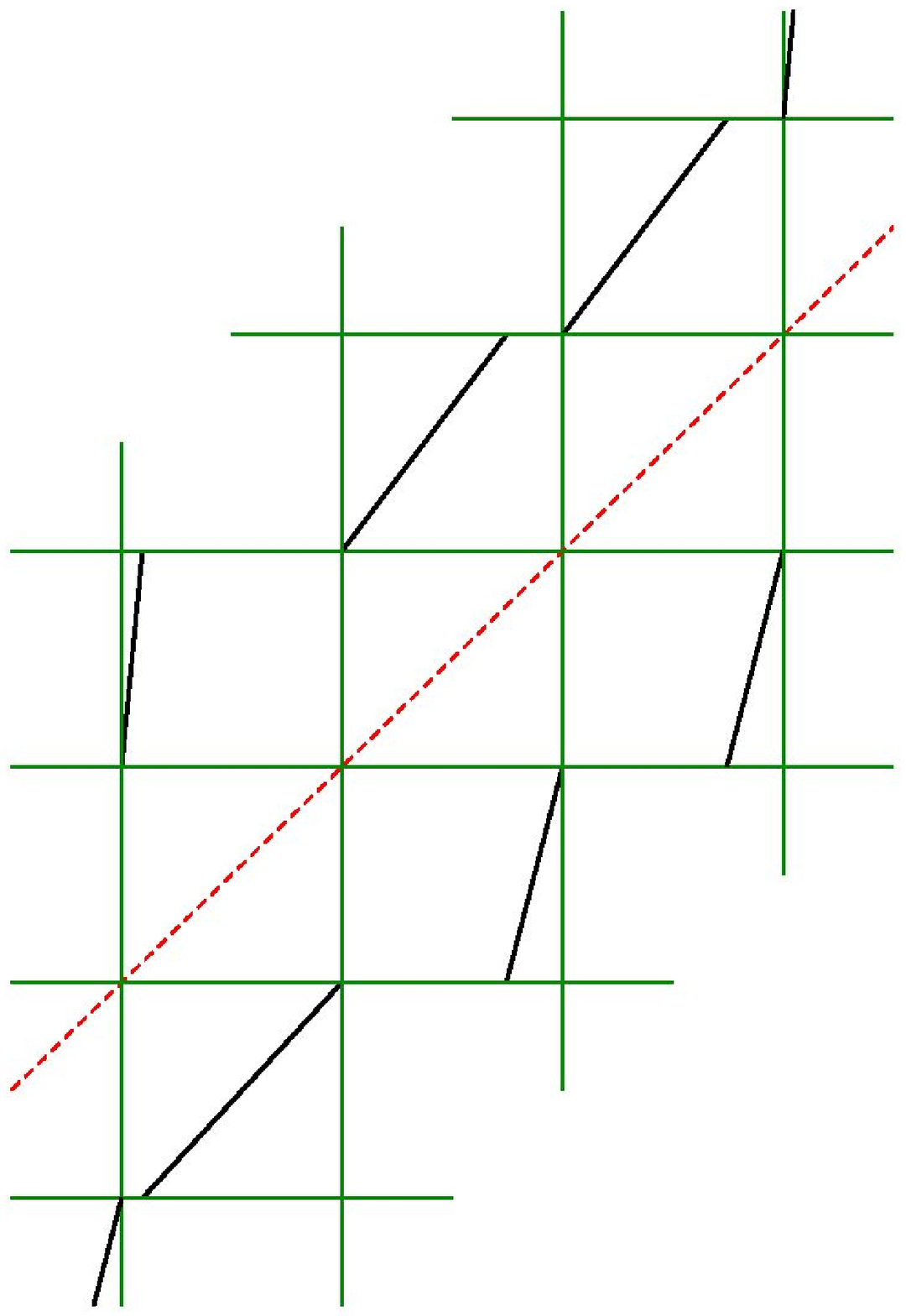}}
\vskip4mm
\caption{\label{Breal}The map corresponding to game $\mathcal{B}$}
\end{figure}

\begin{figure}[h!]
%\label{normvsint}
\centerline{\epsfxsize=8.cm \epsfbox{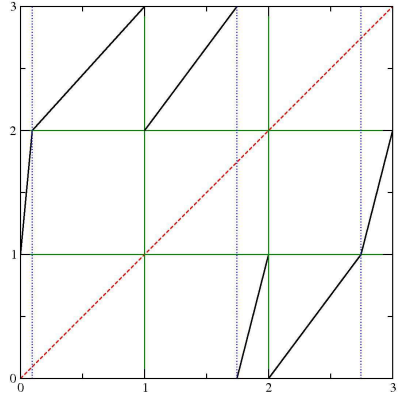}}
\vskip4mm
\caption{\label{Btorus}The torus map corresponding to game $\mathcal{B}$}
\end{figure}

The construction of a dynamical mapping corresponding to successive playing at $\mathcal{B}$ game is slightly more complicated, as the natural unit is not a single cell, but a group of three cells, due to the $mod \, 3$ condition: once we take it into account we may define in this case
\begin{equation}
F_{P\bgame}(x)=\left\{
\begin{array}{ll}
\frac{1}{p_1}x+1 & \mbox{   } x\in[0,p_1) \\
\frac{1}{1-p_1}x-\frac{1}{1-p_1} & \mbox{   } x\in[p_1,1) \\
\frac{1}{p_2}x-\frac{1}{p_2}+2 & \mbox{   } x\in[1,1+p_2) \\
\frac{1}{1-p_2}x-\frac{2}{1-p_2}+1 & \mbox{   } x\in[1+p_2,2) \\
\frac{1}{p_2}x-\frac{2}{p_2}+3 & \mbox{   } x\in[2,2+p_2) \\
\frac{1}{1-p_2}x-\frac{3}{1-p_2}+2 & \mbox{   } x\in[2+p_2,3).
\end{array}
\right.
\label{FPB}
\end{equation}
Fig. (\ref{Breal}) and (\ref{Btorus}) show the corresponding map on the real line
and on the ($mod \, 3$ torus), respectively.
Now we are interested in a random combination of the two games: we call $\gamma$ the probability that at each step game $\mathcal{A}$ is played ($1-\gamma$ will correspondingly be the probability of playing game $\mathcal{B}$): though a treatment involving composition of the individual games' transfer operators is possible \cite{Dab}, the simplest way of constructing the relevant dynamical map is to consider combined rates of winning and losing, and then designing the corresponding map. We define
\beq
q_1=\gamma p +(1-\gamma) p_1
\label{ranw1}
\eeq
and
\beq
q_2=\gamma p +(1-\gamma)p_2
\label{ranw2}
\eeq
these are the rates of winning if the capital is/is not equal to unity ($mod \, 3$):
the corresponding dynamical map is then very similar to (\ref{FPB}) and precisely
\begin{equation}
F_{P\cgame}(x)=\left\{
\begin{array}{ll}
\frac{1}{q_1}x+1 & \mbox{   } x\in[0,q_1) \\
\frac{1}{1-q_1}x-\frac{1}{1-q_1} & \mbox{   } x\in[q_1,1) \\
\frac{1}{q_2}x-\frac{1}{q_2}+2 & \mbox{   } x\in[1,1+q_2) \\
\frac{1}{1-q_2}x-\frac{2}{1-q_2}+1 & \mbox{   } x\in[1+q_2,2) \\
\frac{1}{q_2}x-\frac{2}{q_2}+3 & \mbox{   } x\in[2,2+q_2) \\
\frac{1}{1-q_2}x-\frac{3}{1-q_2}+2 & \mbox{   } x\in[2+q_2,3).
\end{array}
\right.
\label{FPcomb}
\end{equation}
\begin{figure}[h!]
\centerline{\epsfxsize=9.cm \epsfbox{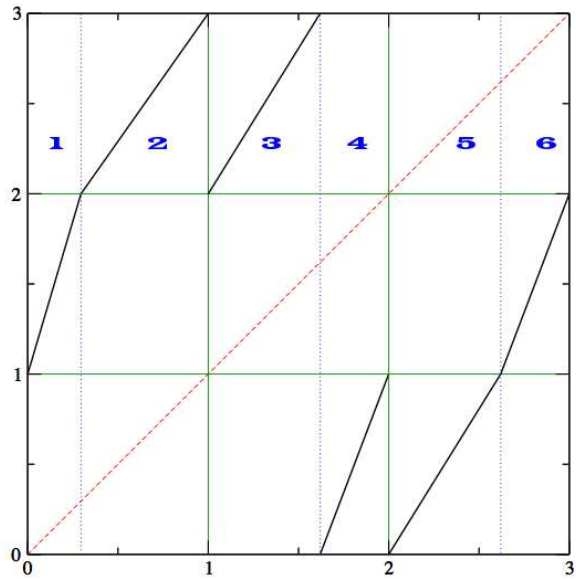}}
\vskip4mm
\caption{\label{Rtorus}The torus map corresponding to the random composition of games $\mathcal{A}$ and $\mathcal{B}$}
\end{figure}
The (torus) map is shown in Fig. (\ref{Rtorus}): the labels $\eta \in \{1,\, 2,\, 3,\, 4,\, 5,\, 6 \}$ indicated on top of the figure are symbols defining a partition of the unit interval according to the support of the different branches of the map (cfr (\ref{FPcomb})), which we label by $\ell_{\eta}$. It is important to remark that in each subinterval $\eta$ both the slope of the map and the capital gain $\sigma_{\eta}$ are
constant: in particular $\sigma_1=\sigma_3=\sigma_5=+1$ (winning intervals) and $\sigma_2=\sigma_4=\sigma_6=-1$ (losing intervals). The slopes (which account for orbit instability) of the different branches are
\beq
\Lambda_3=\Lambda_5=q_2^{-1}\,\,\,\,\,\,\Lambda_4=\Lambda_6=(1-q_2)^{-1}\,\,\,\,\,\,\Lambda_1=q_1^{-1}\,\,\,\,\,\,\Lambda_2=(1-q_1)^{-1}
\label{PFslope}
\eeq
The asymptotic properties of the combined game in this dynamical context correspond to transport features: we are particularly interested in evaluating the net directed velocity $v$
\beq
\Sigma_1(n)\,=\,\langle X_n -X_0 \rangle_0 \,=\, v\cdot n
\label{current}
\eeq
and the diffusion coefficient $D$
\beq
\Sigma_2(n)\,=\,\langle (X_n -X_0 )^2 \rangle_0\,-v^2n^2=\,2D\cdot n
\label{variance}
\eeq
where $\langle \dots \rangle_0$ denotes an average over a set of initial conditions
(for instance uniform distribution on the cell at the origin. 
\section{Periodic orbit theory}
A suitable technique to compute exponents like $v$ and $D$ is via {\em cycle expansions} \cite{Webbook,AAC}: it automatically encompasses universality features, being invariant under topological conjugacies and, while allowing to get analytic results in the present context, it may be applied -as a genuine perturbative scheme- to possible generalizations (for instance if we consider maps where piecewise linearity is lost). Such a technique has been applied to other systems exhibiting chaotic transport \cite{zdiff}, and also generalized to yield anomalous moments' exponents when intermittency appears \cite{AC}. We very briefly recall how $v$ and $D$ are computed: a pedagogical introduction is contained in \cite{Alnp}. The asymptotic behavior of moments is determined through the generating function
\begin{equation}\label{genfunc}
\mathrm{G}_n(\beta) =\quad \langle \rme^{\beta (X_n-X_0)} \rangle
\end{equation}
as
\begin{equation}
\langle(X_n-X_0)^q \rangle= \left.\left(\frac{\partial^q}{\partial \beta^q}   \mathrm{G}_n(\beta)\right) \right|_{\beta=0}
\end{equation}
In this way one has to characterize the asymptotic behavior of the generating function: as in statistical mechanics of lattice systems a (generalized) transfer operator $\mathcal{L}_{\beta}$ may be introduced, such that 
\beq\label{tr}
  \mathrm{G}_n(\beta)=\int{\rmd x [\mathcal{L}^n_{\beta}\rho_{\rm in}](x)}
\eeq
where $\rho_{{\rm in}}$ is the density of initial conditions: thus the leading eigenvalue $\lambda(\beta)$ of the transfer operator dominates the asymptotic behavior of the generating function:
\beq\label{leading}
\mathrm{G}_n(\beta)\,\sim\, \lambda(\beta)^n
\eeq
while
the expression for the transfer operator is as follows:
\begin{equation}\label{transf}
[\mathcal{L}_{\beta}\phi](x)=\int{\rmd y \, \,\rme^{\beta (F_P(y)-y)}\delta(F_P(y)-x)\phi(y)}
\end{equation}
A convenient way to compute the leading eigenvalue of $\mathcal{L}_{\beta}$ is provided by the {\em dynamical zeta function} \cite{Webbook,AAC} which is expressed
as an infinite product over prime periodic orbits of the torus map $\hat{F}$:
\beq\label{dzf}
\zeta^{-1}_{(0)\beta}(z) \,= \,\prod_{\{p\}}\left( 1-\frac{z^{ n_p} \rme^{\beta \sigma_p}}{|\Lambda_p| } \right)
\eeq
each orbit carries a weight determined by prime period $n_p$,  instability $\Lambda_p=\prod_{i=0}^{n_p-1}\hat{F}_P'(\hat{F}_P^i(x_p))$ and the capital gain $\sigma_p$, that is the integer factor that gives the capital variation per cycle once the orbit is unfolded on the real line: if the orbit has symbol sequence
$\epsilon_1,\, \dots \, \epsilon_{n_p}$ according to the partition defined earlier, then
\beq\label{capgain}
\sigma_p\,=\,\sum_{k=1}^{n_p}\,\sigma_{\epsilon_k}
\eeq
The way dynamical zeta functions enter the game is that their smallest zero coincides with the inverse of the leading eigenvalue of the transfer operator \cite{Webbook}: though this might look as a cumbersome way to attack the problem of averages computations, it turns out that if the symbolic dynamics of the system is under control (as in the present example, as we will show later) dynamical zeta functions may provide exact results (as in the Parrondo case), or, in more general examples, they yield a scheme to compute averages in a rapidly converging perturbative scheme \cite{AAC,Webbook}. We call $z(\beta)$ the smallest zero of the dynamical zeta function: by expanding (\ref{genfunc},\ref{leading}) around $\beta=0$ we get
\begin{eqnarray} \label{dev}
\langle e^{\beta (X_n-X_0)}\rangle & \sim & \langle 1+ \beta(X_n-X_0)+\frac{\beta^2}{2}(X_n-X_0)^2+\ldots\rangle\\
\wr \;\;\;\; & & \nonumber\\
\lambda(\beta)^n &\sim & 1+\beta n \lambda'(0)+\frac{\beta^2}{2}\left[n(n-1)\lambda'(0)^2+n\lambda''(0)\right]+\ldots.\nonumber
\end{eqnarray}
where we have taken into account that $\lambda(0)=z(0)^{-1}=1$, as the generalized transfer
operator for $\beta=0$ coincides with the Perron-Frobenius operator whose leading eigenvalue (corresponding to the invariant measure) is $\lambda=1$. 
This easily yields $v$ and $D$ in terms of zeta functions \cite{zdiff} (notice that all cases dealt with in former papers however consider systems where symmetry yields $v=0$)
\begin{equation} \label{vzeta}
v=\frac{d\langle X_n-X_0\rangle}{dn}=\lambda'(0)=\left. -\left(-\frac{\partial_{\beta}\zeta^{-1}_{(0)\beta}}{\partial_z\zeta^{-1}_{(0)\beta}}\right)\right|_{\beta=0,z=1}.
\end{equation}
 For the diffusion constant we get
\begin{equation}
D=\frac{1}{2}\left(\lambda''(0)-\lambda'(0)^2\right).
\end{equation}
where the first derivative is given by (\ref{vzeta}), while
\begin{eqnarray}\label{Dzeta}
\hspace{-1cm}\lambda''(0)
= \left. 2\left(-\frac{\partial_{\beta}\zeta^{-1}_{(0)\beta}}{\partial_z\zeta^{-1}_{(0)\beta}}\right)^2\right|_{z=1\,\beta=0}+\nonumber\\
\hspace{-1cm} + \left. \left( \frac{
 \left({\partial_z\zeta^{-1}_{(0)\beta}}\right)^2{\partial_{\beta \beta} \zeta^{-1}_{(0)\beta}}+
 \left({\partial_{\beta}\zeta^{-1}_{(0)\beta}}\right)^2
 {\partial_{zz}\zeta^{-1}_{(0)\beta}}-2{\partial_{\beta}\zeta^{-1}_{(0)\beta}} {\partial_z\zeta^{-1}_{(0)\beta}}{\partial_{z\beta}\zeta^{-1}_{(0)\beta}}}{\left(\partial_z\zeta^{-1}_{(0)\beta} \right)^3}\right)\right|_{z=1\,\beta=0} 
\end{eqnarray}
\section{Parrondo transport from zeta functions}
\begin{figure}[h!]
\centerline{\epsfxsize=6.cm \epsfbox{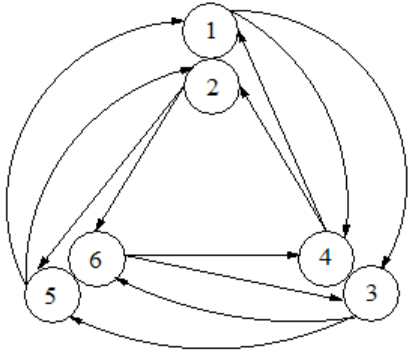}}
\vskip4mm
\caption{\label{Mgraph}The Markov graph generating symbol sequences for the map
$\hat{F}_{P\cgame}$}
\end{figure}

\begin{figure}[h!]
\vskip1cm
\centerline{\epsfxsize=10.cm \epsfbox{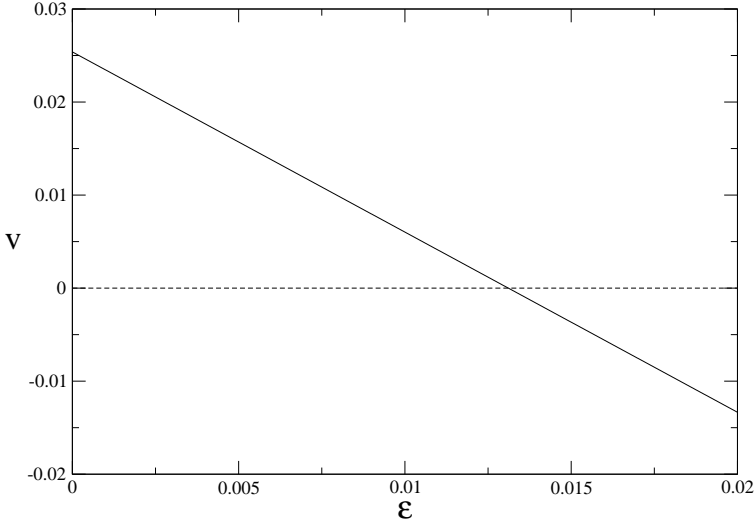}}
\vskip4mm
\caption{\label{ppar}The current $v$ as a function of "unfairness" parameter $\epsilon$ for small positive $\epsilon$ ($\gamma=1/2$)}
\end{figure}

\begin{figure}[h!]
\vskip1cm
\centerline{\epsfxsize=10.cm \epsfbox{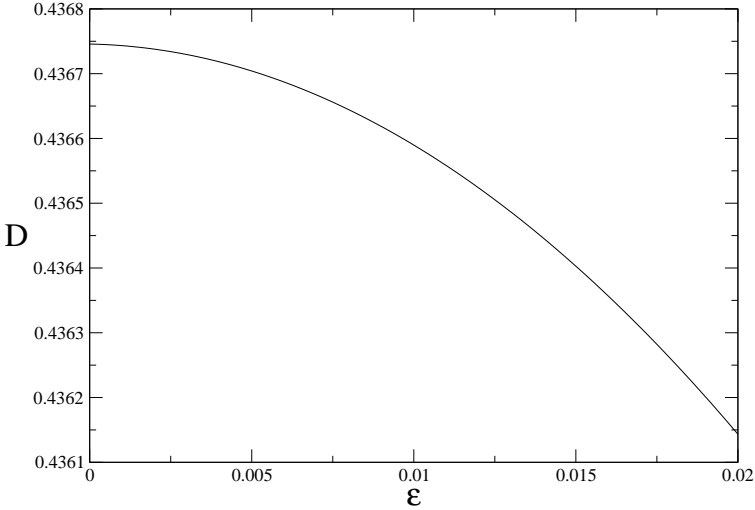}}
\vskip4mm
\caption{\label{Dpar}The diffusion coefficient $D$ as a function of "unfairness" parameter $\epsilon$ ($\gamma=1/2$)}
\end{figure}

\begin{figure}[h!]
\vskip1cm
\centerline{\epsfxsize=10.cm \epsfbox{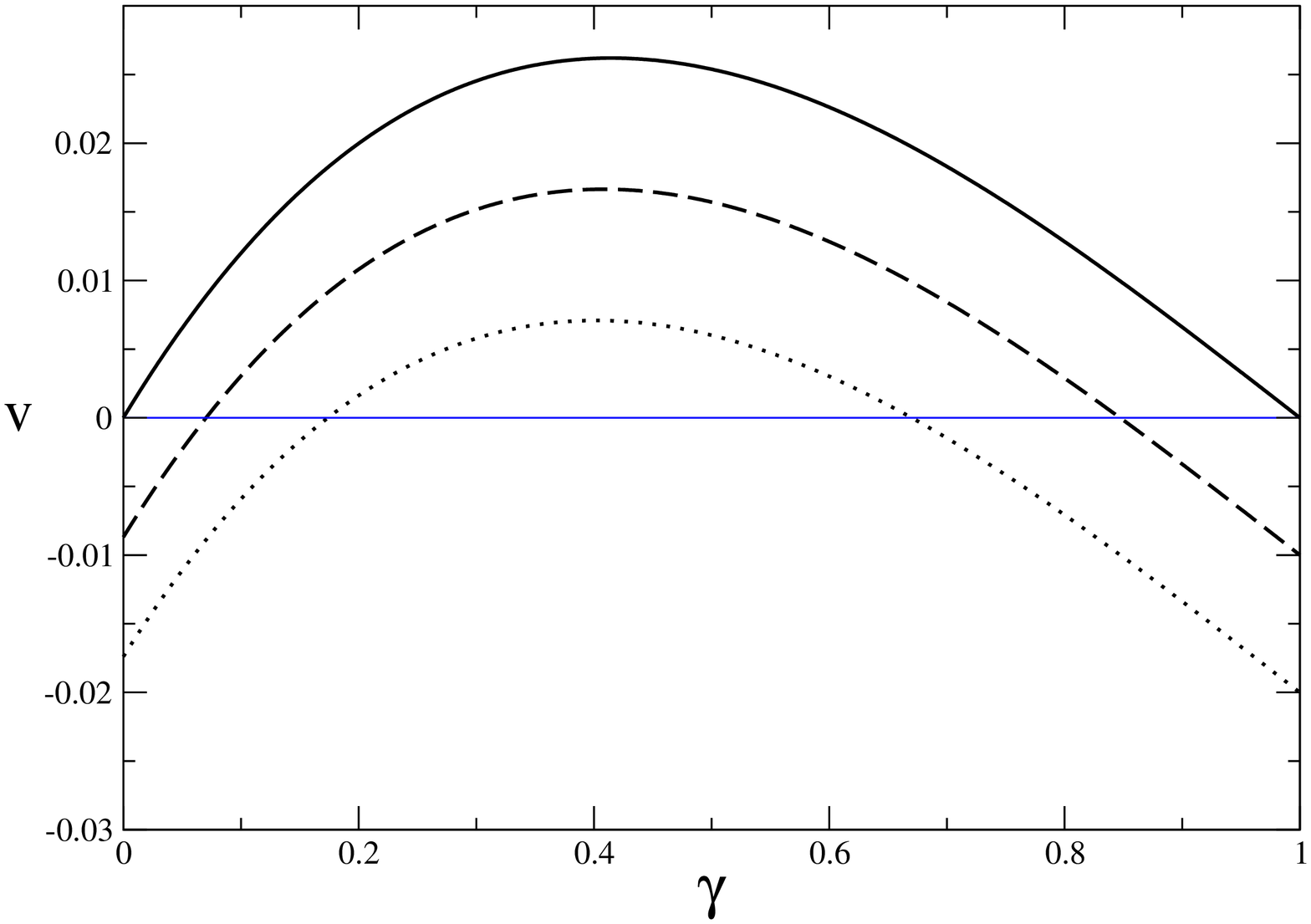}}
\vskip4mm
\caption{\label{cug}The current $v$ as a function of $\gamma$ for $\epsilon=0$ (full line), $\epsilon=0.005$ (dashed line) and $\epsilon=0.01$ (dotted line). Maxima are attained at $\gamma=0.414588$, $\gamma=0.408161$ and $\gamma=0.401643$, respectively.}
\end{figure}

In order to use formulas like (\ref{vzeta},\ref{Dzeta}) for the deterministic version of Parrondo games we have to provide an expression for the zeta function. This is done once we set up the symbolic rules for dynamics: by considering the torus map $\hat{F}_{P\cgame}$ (see fig. (\ref{Rtorus})) the only allowed symbolic transitions are the following 
\beq\label{symball}
\begin{array}{llll}
...13... & ...14... & ...25... & ...26... \\
...35... & ...36... & ...41... & ...42... \\
...51... & ...52... & ...63... & ...64... \end{array}
\eeq
thus admissible symbolic sequences are generated by the Markov graph of fig. (\ref{Mgraph})
As the weights we attach to cycles depend only on how many times different symbols appear in the cycle code, our problem is essentially topological and the zeta functions is a polynomial, which we can easily write down once we have the full list of non-intersecting loops of the Markov graph of fig. (\ref{Mgraph}). These are 
\beq\label{loops}
\overline{14}\,\,\,\,\overline{25}\,\,\,\,\overline{36}\,\,\,\,\overline{135}\,\,\,\,\overline{264}\,\,\,\,\overline{1364}\,\,\,\,\overline{1425}\,\,\,\,\overline{2635}\,\,\,\,\overline{135264}\,\,\,\,\overline{136425}\,\,\,\,\overline{142635}
\eeq
Then \cite{topMG,gra}
\beq\label{topexp}
\zeta^{-1}_{(0)\beta}\,=\,\sum_{k=0}^f\,\sum_{\pi}\,(-1)^k\,t_{p_1}\cdots t_{p_k}
\eeq
where all possible combinations $\pi$ of non intersecting loops $p_1 \cdots p_k$ are considered, $f$ is the maximum number of non intersecting loops that we may put on
the graph and
\beq\label{gent}
t_p\,=\,\frac{z^{n_p}e^{\beta \sigma_p}}{\Lambda_p}
\eeq
Once we take into account cancellations (for instance the cycle $\overline{142635}$ is completely shadowed by the product of $\overline{14}$ and 
$\overline{2635}$ cycles) the generalized zeta function is written as
\begin{eqnarray}\label{Pzeta}
\zeta^{-1}_{(0)\beta}(z)&\,=\,&1-z^2\left(\frac{e^{\sigma_{14}}}{\Lambda_{14}}+\frac{e^{\sigma_{25}}}{\Lambda_{25}}+\frac{e^{\sigma_{36}}}{\Lambda_{36}}\right)-z^3\left(\frac{e^{\sigma_{135}}}{\Lambda_{135}}+\frac{e^{\sigma_{264}}}{\Lambda_{264}} \right) \\ &\,=\,& 1-z^2\left(q_1(1-q_2)+q_2(1-q_1)+q_2(1-q_2)\right)-\nonumber \\ & \, & +z^3 \left( 
e^{3\beta}q_1q_2^2+e^{-3\beta}(1-q_1)(1-q_2)^2 \right) \nonumber
\end{eqnarray}
where we have taken into account how instabilities (\ref{PFslope}) are expressed in terms of transition probabilities (\ref{ranw1},\ref{ranw2}).
Now we take the original Parrondo values 
\beq\label{Ppar}
p\,=\,\frac{1}{2}-\epsilon \quad p_1\,=\,\frac{1}{10} - \epsilon \quad p_2 \,= \, \frac{3}{4}-\epsilon
\eeq
and take $\gamma=1/2$: once we insert (\ref{Pzeta}) into (\ref{vzeta}) we may verify Parrondo paradox, namely how unfair games (for small positive $\epsilon$) lead to capital gain: see fig. (\ref{ppar}), and in the same way we may look at the behavior of the diffusion coefficient (see fig. (\ref{Dpar}))
In the same fashion we may check how transport exponents depend on the parameter $\gamma$ once the bias $\epsilon$ is fixed (see the current behavior in fig. (\ref{cug}))

It turns out that by using the explicit form of the zeta function we can actually derive analytical expressions for such quantities: for instance the current $v$ as a function of $\epsilon$ for $\gamma=1/2$ reads (in agreement with \cite{Parr2})
\beq\label{curr05}
v_{1/2}(\epsilon)\,=\,\frac{6\left(3-229\epsilon +16 \epsilon^2 -320\epsilon^3\right)}{709-32\epsilon +960\epsilon^2}
\eeq
while, more generally we can express the current $v$ as a function of both $\epsilon$ and $\gamma$ as 
\beq\label{currv}
\hspace{-1cm}v_{\gamma}(\epsilon)\,=\,\frac{6 \left(-80 \epsilon ^3+8 (1-\gamma )
   \epsilon ^2-(11 (2-\gamma ) \gamma +49) \epsilon
   +2 (1-\gamma ) (2-\gamma ) \gamma \right)}{240
   \epsilon ^2-16 (1-\gamma ) \epsilon +11
   (2-\gamma ) \gamma +169}
\eeq
%% mod 20-10 {
Analytic expression may be obtained also for the diffusion constant $D$, for instance
\begin{eqnarray}\label{Dhalf}
\hspace{-1.8cm}D_{1/2}(\epsilon)\,=\,-(9 (4 \epsilon  (\epsilon  (64 \epsilon 
   (\epsilon  (64 \epsilon  (5 \epsilon  (320
   \epsilon  (15 \epsilon
   -2)+9243)-4487)+1458885)+\nonumber\\
   \hspace{-0.8cm}-69761)-6333811)+1191650
   )-34590345))/(2 (32 \epsilon  (30 \epsilon
   -1)+709)^3)
\end{eqnarray}
%% mod 20-10 }
%
\section{Conclusions}
We have constructed a one-dimensional mapping providing a deterministic version of Parrondo games: once the symbolic dynamics is coded, periodic orbit theory offers a way to get analytic estimates of all the relevant transport exponents. 
While in this paradigmatic example results may be obtained via probabilistic techniques
\cite{Parr2,ACC} we remark two things: once the structure of the relevant zeta function is explicitly derived all statistical averages may be computed in a similar way in the present framework: moreover our work may be considered as a first step towards the periodic orbit analysis of ``nonlinear" Parrondo games, where the map ceases to be piecewise linear: in particular we plan to investigate sticking effects in this context, by considering intermittent Parrondo maps. 
\ack	
This work was partially supported by the PRIN 2003 Project {\em Order and Chaos in Nonlinear Extended Systems}.

%\end{multicols}
\end{document}